\documentclass[twocolumn,floatfix]{revtex4}
\usepackage{graphicx}
\usepackage{dcolumn}
\usepackage{amsmath}
\newcommand{\Bo}{{B}_{\rm 0}}
\newcommand{\dBpa}{\Delta B\mathrm{_{n}^{\parallel}}}
\newcommand{\dBpe}{\Delta B\mathrm{_{n}^{\perp}}}

\newcommand{\Bpa}{B\mathrm{_{n}^{\parallel}}}
\newcommand{\Bpe}{B\mathrm{_{n}^{\perp}}}
\newcommand{\Btot}{B_\mathrm{tot}}

\hsize\textwidth\columnwidth\hsize\csname@twocolumnfalse
\endcsname
\newcounter{lastnote}

\begin{document}
\title{Suppressing Spin Qubit Dephasing by Nuclear State Preparation}

\author
{D. J. Reilly$^{1}$, J. M. Taylor$^{2}$, J. R. Petta$^3$, C. M. Marcus$^1$\\
 M. P. Hanson$^4$ and A. C. Gossard$^4$\\
\normalsize{$^1$ Department of Physics, Harvard University, Cambridge, MA 02138, USA}\\
\normalsize{$^2$ Department of Physics, Massachusetts Institute of Technology, Cambridge, MA 02139, USA}\\
\normalsize{$^3$ Department of Physics, Princeton University, Princeton, NJ 08544, USA}\\
\normalsize{$^4$ Materials Department, University of California, Santa Barbara, California 93106, USA} 
\\}

\begin{abstract}
Coherent spin states in semiconductor quantum dots offer promise as electrically controllable quantum bits (qubits) with scalable fabrication. For few-electron quantum dots made from gallium arsenide (GaAs), fluctuating nuclear spins in the host lattice are the dominant source of spin decoherence. We report a method of preparing the nuclear spin environment that suppresses the relevant component of nuclear spin fluctuations below its 
equilibrium value by a factor of $\sim70$, extending the inhomogeneous dephasing time for the two-electron spin state beyond 1 $ \mu s$. The nuclear state can be readily prepared by electrical gate manipulation and persists for $>$ 10~s. 
  \end{abstract}
\maketitle

\begin{figure}[t!!]
\begin{center}
\includegraphics[width=8.0cm]{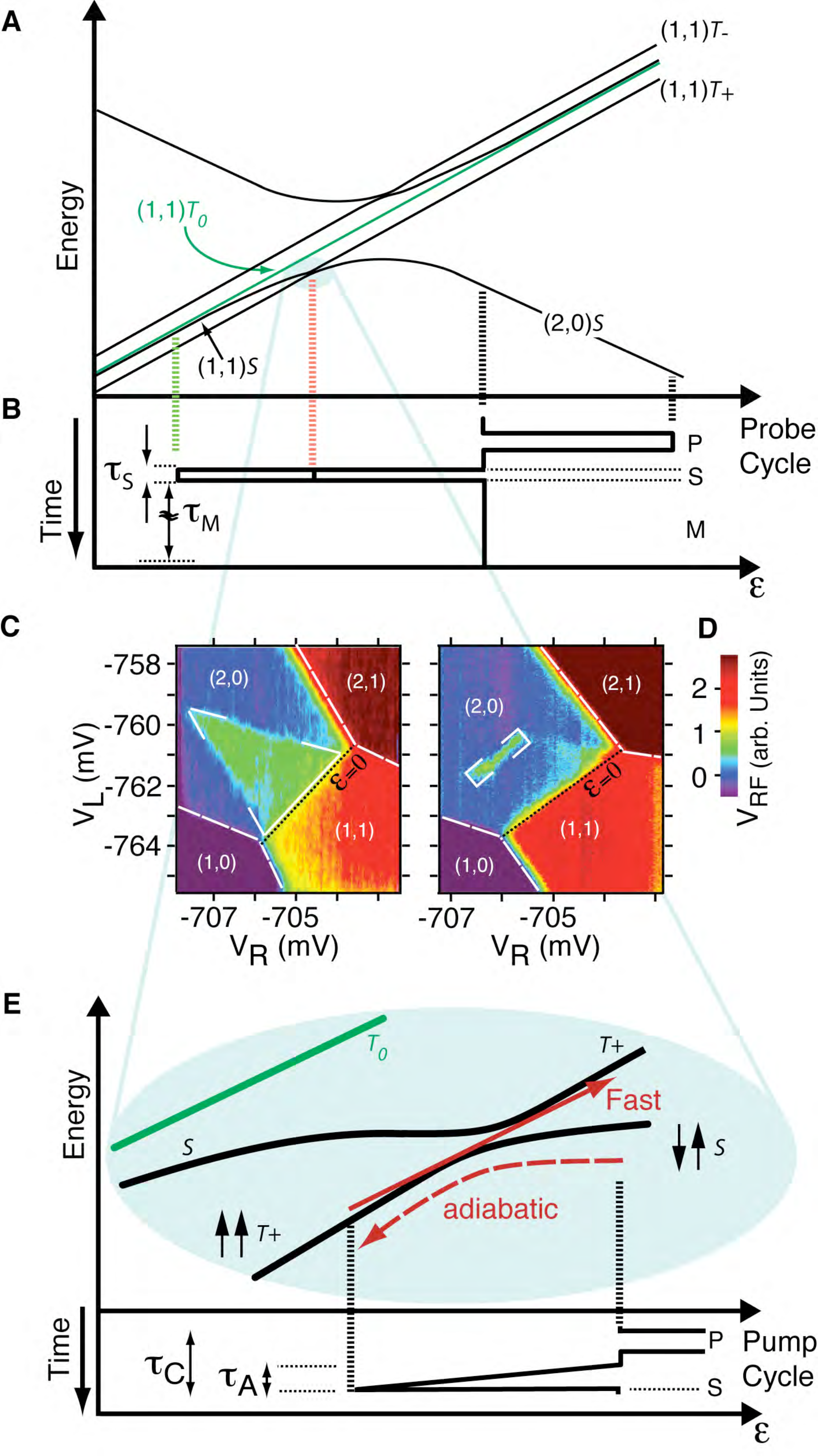}
\caption{({\bf{A}}) Schematic of the energy levels of the two-electron system
  in a magnetic field. Detuning, $\varepsilon$, from the (2,0)-(1,1) charge
  degeneracy is gate controlled. ({\bf{B}}) Gate-pulse sequence used to separately
  probe the longitudinal, $\dBpa$ (green dashed line), and transverse, $\dBpe$ (red dashed line) components of the Overhauser field difference, depending on the position of the separation point S. ({\bf{C}} and {\bf{D}}) Time-averaged charge-sensing signal, $V_{\rm
    rf}$ from the rf-QPC, as a function of gate biases $V_{L}$ and $V_{R}$, showing
  features corresponding to the singlet mixing with the $T_{0}$
  (bracketed green triangle in C) and $T_{+}$ (bracketed green line segment in D). ({\bf{E}}) Schematic view of the $S-T_{+}$
  anticrossing, illustrating the pumping cycle. With each iteration of
  this cycle, with period $\tau_C = 250$ ns, a new singlet state is taken adiabatically through the
  $S-T_{+}$ anticrossing in a time $\tau_A = 50$ ns, then returned nonadiabatically to (2,0) in $\sim 1$ ns, where the $S$ state is then reloaded. }
\vspace{-0.5cm}
\end{center}
\end{figure}

Quantum information processing requires the realization of interconnected,
controllable quantum two-level systems (qubits) that are sufficiently isolated from their environment that quantum coherence can be maintained for much longer than the characteristic operation time. Electron spins in quantum dots are an appealing candidate system for this application, as the spin of the electron is typically only weakly coupled to the environment compared to the charge degree of freedom~\cite{loss98}. Logical qubits formed from pairs of spins provide additional immunity from collective dephasing, forming a dynamical decoherence-free subspace~\cite{levy02,taylor05}. 

Implementing any spin-qubit architecture requires the manipulation~\cite{petta05,koppens06,nowack07} and detection~\cite{elzerman04,amasha08} of few-electron spin states, as yet only demonstrated in III-V semiconductor heterostructure devices such as gallium arsenide (GaAs), which, in all cases, comprise atoms with nonzero nuclear spin. The nuclear spins of the host lattice couple to electrons via the hyperfine interaction and causes rapid electron spin dephasing. In the GaAs devices presented here, for instance, an ensemble of initialized spin pairs will retain their phase relationship for $T_{2}^{*} \sim$~15~ns, consistent with theoretical estimates \cite{erlingsson01,khaetskii02,merkulov02} and previous measurements \cite{petta05}. The time $T_{2}^{*}$ represents an inhomogeneous dephasing time, which can be extended using spin-echo methods \cite{petta05}. Extending $T_{2}^{*}$ by nuclear state preparation reduces the burden of using echo sequences or large field gradients to overcome the fluctuating hyperfine fields when controlling spin qubits.

Proposals to reduce dephasing by nuclear state preparation include complete nuclear polarization~\cite{imamoglu03}, state-narrowing of the nuclear distribution~\cite{imamoglu03,klauser06,giedke06,stepanenko06}, and
schemes for decoupling the bath dynamics from the coherent evolution
of the electron spin using control pulses~\cite{khodjasteh05,zhang07,yao07}. These approaches remain largely
unexplored experimentally, though recent optical experiments~\cite{greilich07} have demonstrated a suppression of nuclear fluctuations in ensembles of self-assembled quantum dots.

\begin{figure*}[t!!]
\begin{center}
\includegraphics[width=18.0cm]{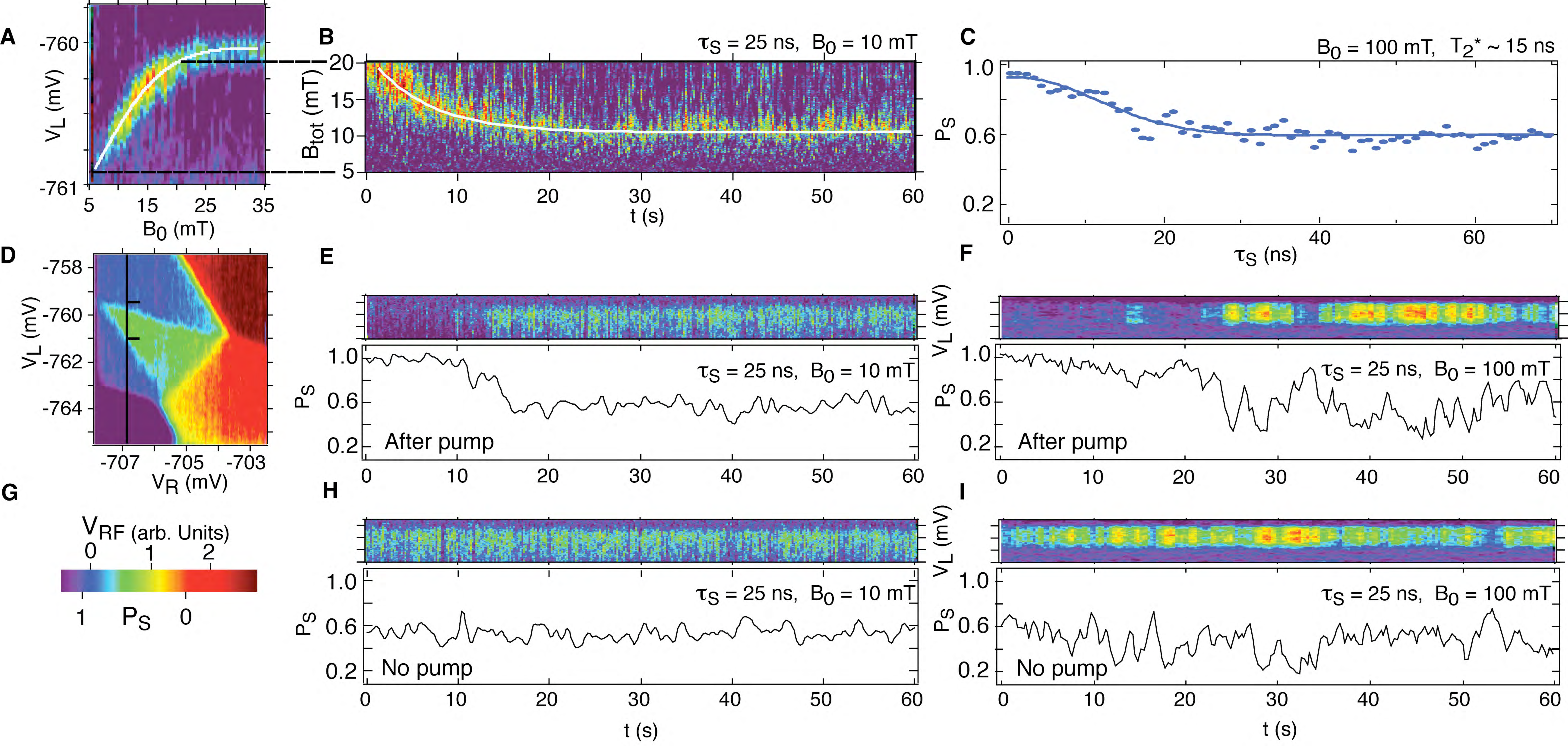}
\caption{
\noindent ({\bf{A}}) Position of the $S-T_{+}$ resonance in $V_{L}$ as a function of applied magnetic field amplitude $B_\mathrm{0}$, without prior pump cycle.  ({\bf{B}}) Evolution of the $S-T_{+}$ position as a function of time following the pump cycle. Resonance position in gate voltage $V_{L}$ and converted to $B_\mathrm{tot}$ via  (A). ({\bf{C}}) Singlet return probability $P_{S}$ as a function of $\tau_{\rm S}$ at $B_\mathrm{0}$ = 100 mT . Gaussian fit gives an inhomogeneous dephasing time $T_{2}^{*}$ = 15 ns. ({\bf{D}}) Sensor output $V_{\rm rf}$, as in Fig.~1C, showing triangle that yields $P_{S}$. Vertical cut (black line) with brackets shows location of slices in upper panels of (E-I). ({\bf{E-F}}) Following the pump cycle, repeated slices across this triangle at the position indicated by the black line in (D) allow a calibrated measure of $P_{S}$ as it evolves in time, at (E) $B_\mathrm{0}$ = 10 mT and (F) $B_\mathrm{0}$ = 100 mT. ({\bf{G}}) shows the calibration between the sensing signal $V_{\rm rf}$ and singlet return probability $P_{S}$.  ({\bf{H}} and {\bf{I}}) are control experiments showing slices across the triangle as in (E) and (F), but without a prior pump cycle.} 
\vspace{-0.5cm}
\end{center}
\end{figure*}

We demonstrate a nuclear state preparation scheme in a double quantum-dot system using an electron-nuclear flip-flop pumping cycle controlled by voltages applied to electrostatic gates. Cyclic evolution of the two-electron state through the resonance between the singlet ($S$) and $m_{s} = 1$ triplet ($T_{+}$)~\cite{petta08} leads to a 70-fold suppression of fluctuations below thermal equilibrium of the hyperfine field gradient between the dots along the total field direction.  It is this component of the hyperfine field gradient that is responsible for dephasing of the two-electron spin qubit formed by $S$ and $m_s = 0$ triplet ($T_0$) states \cite{petta05}. Consequently, though the flip-flop cycle generates only a modest nuclear polarization ($<$ 1\%), the resulting nuclear state extends $T_{2}^{*}$ of the $S-T_0$ qubit from 15~ns to beyond 1~$\mu$s.  Once prepared, this non-equilibrium nuclear state persists for $\sim$15 s, eventually recovering equilibrium fluctuations on the same time scale as the relaxation of the small induced nuclear polarization. Noting that this recovery time is $\sim 9-10$ orders of magnitude longer than typical gate operation times, we propose that occasional  nuclear state preparation by these methods may provide a remedy to hyperfine-mediated spin dephasing in networks of interconnected spin qubits.

The double quantum dot is defined in a GaAs/AlGaAs heterostructure with a two-dimensional electron gas (2DEG) 100 nm below the wafer surface (density 2$\times$10$^{15}\, {\rm
  m}^{-2}$, mobility 20 m$^{2}$/Vs). Negative voltages applied to
Ti/Au gates create a tunable double-well potential that is
tunnel coupled to adjacent electron reservoirs [see Fig.~S1 of Supporting Online Material (SOM)]. A proximal radio-frequency quantum point contact (rf-QPC) senses the charge state of the double dot, measured in terms of the rectified sensor output voltage  $V_\mathrm{rf}$ \cite{reilly07}. Measurements were made in a dilution refrigerator at base electron temperature of 120 mK.

A schematic energy level diagram (Fig.~1A), with (n,m) indicating equilibrium charge occupancies of the left and right dots, shows the three (1,1) triplet states ($T_{+}$, $T_{0}$, $T_{-}$) split by a magnetic field $\bf{\Bo}$ applied perpendicular to the 2DEG. The detuning  from the (2,0)-(1,1)
degeneracy, $\varepsilon$, is controlled by high-bandwidth gate voltages pulses. The ground state of (2,0) is a singlet, with the (2,0) triplet out of the energy range of the experiment.

Each confined electron interacts with $N \sim 10^{6}$ nuclei via hyperfine coupling, giving
rise to a spatially and temporally fluctuating effective magnetic (Overhauser) field~\cite{erlingsson01,khaetskii02,merkulov02,taylor07}.  In the separated (1,1) state, precession rates for the two electron spins depends on their local effective fields, which can be decomposed into an average field and a difference field. It is useful to resolve  ${\bf{B}_\mathrm{n}} = ({\bf{B}^{l}_\mathrm{n}} +
{\bf{B}^{r}_\mathrm{n}})/2$, the Overhauser part of the total average field,  ${\bf{\Btot}} = {\bf{\Bo}} + {\bf{B}_\mathrm{n}}$, into components  along ($\Bpa$) and transverse ($\Bpe$) to   ${\bf{\Btot}}$. The difference field, due only to  Overhauser contributions, is given by $\Delta{\bf{ B_\mathrm{n}}} = ({\bf{B}^{l}_\mathrm{n}} -
{\bf{B}^{r}_\mathrm{n}})/2$, with components along ($\dBpa$) and transverse
($\dBpe$) to ${\bf{\Btot}}$. At large negative $\varepsilon$, where the two electrons are well separated and exchange $J(\varepsilon)$ is negligible, $\dBpa$ sets the precession rate between $S$ and $T_{0}$ states.  At the value of detuning where $J(\varepsilon)$ equals  the Zeeman energy $E_\mathrm{Z} = g \mu_{B}\Btot$, ($g$ is the electron $g$-factor and $\mu_{B}$ is the Bohr magneton) precession between $S$ and $T_{+}$ states occurs at a rate set by $\dBpe$.

To measure the precession or dephasing of spin pairs in the two dots, a gate-pulse cycle (``probe cycle") first prepares (P) a singlet state in (2,0), then separates (S) the two electrons into (1,1) for a duration $\tau_{\rm S}$, then measures (M) the probability of return to (2,0).  States that evolve into triplets during $\tau_{\rm S}$ remain trapped in (1,1) by the Pauli blockade, and are detected as such by the rf-QPC charge sensor \cite{petta05}. Figures~1C and D show the time-averaged
charge sensing signal, $V_{\rm rf}$, as a function of constant offsets to gate biases
$V_{L}$ and $V_{R}$, with this pulse sequence running continuously.  Setting the amplitude of the S-pulse to mix $S$ with $T_{0}$ at large detuning (green dashed line, Fig.~1B) yields the ``readout triangle" indicated in Fig.~1C. Within the triangle, $V_{\rm rf}$ is between (2,0) and (1,1) sensing values, indicating that for some probe cycles the system becomes Pauli blockaded in (1,1) after evolving to a triplet state. Outside this triangle, alternative spin-independent relaxation pathways circumvent the blockade \cite{johnson04}. For a smaller amplitude S-pulse (red dashed line, Fig.~1B), $S$ mixes with $T_{+}$, also leading to partial Pauli blockade and giving the narrow resonance feature seen in Fig.~1D. The dependence of the $S-T_{+}$ resonance position on applied field $B_0$ serves as a calibration, mapping gate voltage $V_{L}$ (at fixed $V_{R}$) into total effective field, $\Btot$, including possible Overhauser fields  (Fig.~2A). The charge sensing signal, $V_{\rm rf}$, is also calibrated using equilibrium (1,1) and (2,0) sensing values to give the probability $1-P_{S}$ that an initialized singlet will evolve into a triplet during the separation time $\tau_{\rm S}$ (Fig.~2G). A fit to $P_{S}(\tau_{\rm S})$ (Fig.~2C) yields \cite{merkulov02,taylor07,reilly07b} a dephasing time $T^{*}_{2} = \hbar / g \mu_{B} \langle \dBpa \rangle_\mathrm{rms}$ $\sim$ 15 ns, where $\langle ...\rangle_\mathrm{rms}$  denotes a root-mean-square time-ensemble average.

We now investigate effects of the electron-nuclear flip-flop cycle (``pump cycle")  (Fig.~1E). Each iteration of the pump cycle moves a singlet, prepared in (2,0), adiabatically through the
$S-T_+$ resonance, then returns non-adiabatically to (2,0), where the state is re-initialized to a singlet by exchanging an electron with the adjacent reservoir \cite{petta08}. In principle, with each iteration of this cycle, a change in the angular momentum of the electron state occurs with a corresponding change to the nuclear system. Iterating the pump cycle at 4 MHz creates a modest nuclear polarization of order 1\%, as seen previously~\cite{petta08}.  The pump cycle was always iterated for more than 1 s, and no dependence on pumping time beyond 1 s was observed. What limits the efficiency of the pumping cycle, keeping the polarization in the few-mT regime, is not understood.

Immediately following the pump cycle, the gate-voltage pattern is switched to execute one of two types of probe cycles. The first type of probe cycle starts in (2,0) and makes a short excursion into (1,1) to locate the $S-T_{+}$ resonance, allowing $\Btot$ to be measured via Fig.~2A. Figure 2B shows that the nuclear polarization established by the pumping cycle relaxes over $\sim$~15 s. The second type of probe cycle starts in (2,0) and makes a long excursion deep into (1,1) to measure $P_{S}(\tau_{\rm S})$ where exchange is insignificant and $S$ and $T_{0}$ states are mixed by $\dBpa$.   We examine $P_{S}(\tau_{\rm S})$ at fixed $\tau_{\rm S}$ as a function of time after the end of the pump cycle by sampling  $V_\mathrm{rf}$ while rastering $V_{L}$ across the readout triangle. The black line in Fig.~2D shows the value of $V_{R}$ with the tick marks indicating the upper and lower limit of the rastering. Slicing through the readout triangle allows $P_{S}$ to be calibrated within each slice. Remarkably, we find that $P_{S}(\tau_\mathrm{S} = 25$ ns$)$ remains close to unity---that is, the prepared singlet remains in the singlet state after 25 ns of separation---for $\sim$~15~s following the pump cycle (Figs.~2E and 2F). Note that $\tau_\mathrm{S} = 25$ ns exceeds by a factor of $\sim 2$ the value of $T_{2}^{*}$ measured when not preceded by the pump cycle (Fig.~2C). The time after which $P_{S}$ resumes its equilibrium behavior, with characteristic fluctuations \cite{reilly07b} around an average value $P_{S}(\tau_\mathrm{S} = 25$ ns$) = 0.5$, is found to correspond to the time for the small ($\sim 1\%$) nuclear polarization to relax (Fig.~2B). Measurements of $P_{S}(\tau_\mathrm{S} = 25$ ns) using the same probe cycle without the preceding pump cycle, shown in Figs.~2H-I, do not show suppressed mixing of the separated singlet state.

Measurement of $P_{S}(\tau_\mathrm{S})$ as a function of $\tau_\mathrm{S}$ shows that $T_2^*$ for the separated singlet can be extended from 15 ns to 1 $\mu s$, and that this enhancement lasts for several seconds following the pump cycle. These results are summarized in Fig.~3. Over a range of values of $\tau_\mathrm{S}$, slices through the readout triangle (as in Figs.~2E) are sampled as a function of time following pumping, calibrated using the out-of-triangle background, and averaged, giving traces such as those in Figs.~3B-E.  Gaussian fits yield $T^{*}_{2} \sim$ 1 $\mu$s for 0-5 s after the pump cycle and $T^{*}_{2} \sim$ 0.5 $\mu$s for 10-15 s after the pump cycle. After 60 s, no remnant effect of the pump cycle can be seen, with $T^{*}_{2}$ returning to $\sim$ 15 ns, as prior to the pump cycle.
\begin{figure}[t!!]
\begin{center}
\includegraphics[width=8.0cm]{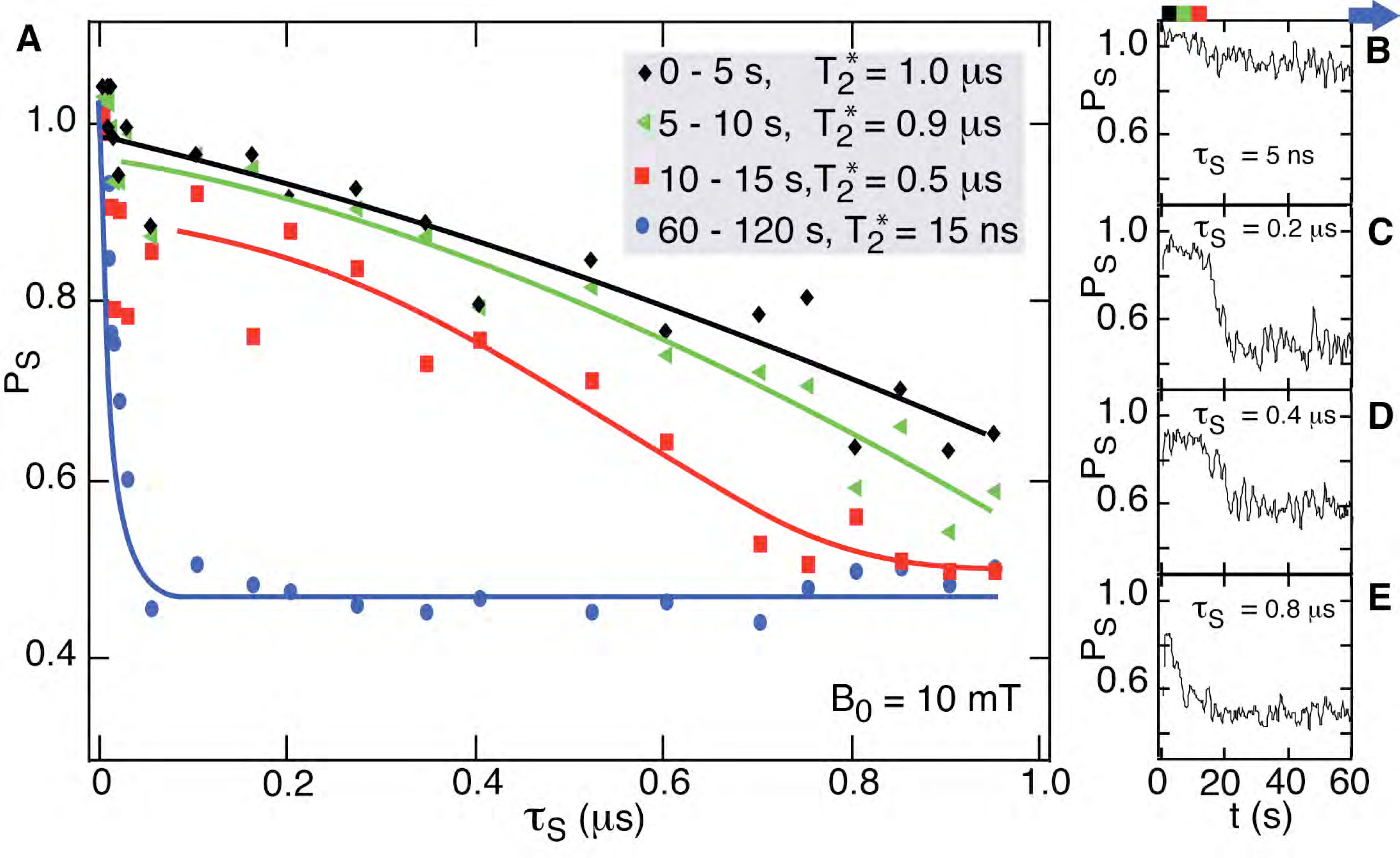}
\caption{
\noindent ({\bf{A}}) Singlet return probability $P_{S}$ as a function of  separation time $\tau_\mathrm{S}$ deep in (1,1) (green dashed line in Fig.~1A), where $S$ mixes with $T_{0}$.  $P_{S}$ averaged within the 0-5 s interval (black), the 5-10 s interval (green), the 10-15 s interval (red), and  60-120 s following the pump cycle, along with gaussian fits. ({\bf{B}}-{\bf{E}}) $P_{S}$ as a function of time following the pump cycle ($B_\mathrm{0}$ = 10 mT) for fixed $\tau_\mathrm{S}$ = 5 ns, $\tau_\mathrm{S}$ = 0.2 $\mu$s,  $\tau_\mathrm{S}$ = 0.4 $\mu$s and $\tau_\mathrm{S}$ = 0.8 $\mu$s.}
\vspace{-0.5cm}
\end{center}
\end{figure}

The rms amplitude of longitudinal Overhauser field difference, $\langle\dBpa\rangle_\mathrm{rms} = \hbar / g \mu_{B} T^{*}_{2}$ is evaluated using $T^{*}_{2}$ values within several time blocks following the pump cycle (using data from Fig.~3A). The observed increase in $T^{*}_{2}$ following the pump cycle is thus recast in terms of a suppression of fluctuations of $\dBpa$ (Fig.~4A).  Similarly,  the $S-T_{+}$ mixing rate is used to infer the size of fluctuations of the transverse component of the Overhauser field, $\langle\dBpe\rangle_\mathrm{rms}$. Figure 4B shows $P_{S}(\tau_\mathrm{S} = 25$ ns$)$ near the $S-T_+$ resonance .  Unlike $S-T_0$ mixing, which is strongly suppressed by the pump cycle, the $S-T_{+}$ resonance appears as strong as before the pump cycle. This suggests that the energy gap $E_\mathrm{n}^\mathrm{\perp}$ (Fig.~4E) is not closed by the pump cycle. Note that fluctuations in $\dBpe$ produce fluctuations in  $E_\mathrm{n}^\mathrm{\perp}$, which give the $S-T_{+}$ anticrossing a width in detuning $\varepsilon$ (Fig.~4E). Converting to a width in magnetic field via Fig.~2A, gives the fluctuation amplitude $\langle\dBpe\rangle_\mathrm{rms}$ following the pump cycle.  Figure 4C shows a representative slice taken from Fig.~4B at the position indicated by the white dashed line. Gaussian fits to each 1 s slice yield mean positions, $m$, and widths, $w$, in magnetic field that fluctuate in time, as seen in Fig.~4D. The increase in $w$ for short times ($t < 10$ s) reflects gate-voltage noise amplified by the saturating conversion from gate voltage to effective field at large $\Btot$ (see SOM).   Beyond these first few seconds, $w$ is dominated by fluctuations of $\dBpe$, but is also sensitive to fluctuations in $m$ that result from fluctuations of $\Bpa$ (see Fig.~4E). (For $t > 10$ s, gate-voltage noise makes a relatively small ($< 10$~\%) contribution to the fluctuations). Estimating and removing the contribution due to $\Bpa$ (see SOM) gives an estimate of $\langle\dBpe\rangle_\mathrm{rms}$ as a function of time following the pump cycle. These results are summarized by comparing Fig.~4A and 4F: in contrast to the strong suppression of fluctuations in $\dBpa$ following the pump cycle, no corresponding suppression of $\langle\dBpe\rangle_\mathrm{rms}$ is observed.

Reducing the cycle rate by a factor $\sim$ 10 reduces, but does not eliminate, the suppression of fluctuations of $\dBpa$ remains evident (see SOM for discussion of dependence of polarization on pump cycle rate). Also, when the pump cycle is substituted by a cycle that rapidly brings the singlet into resonance with $T_{0}$, deep in (1,1), effectively performing multiple fast measurements of $\dBpa$, no subsequent effect on $S-T_0$ mixing is observed. This demonstrates that it is transitions involving $S$ and $T_{+}$, rather than $S$ and $T_{0}$, that lead to the suppression of nuclear field gradient fluctuations.  
\begin{figure}[t!!]
\begin{center}
\includegraphics[width=8.0cm]{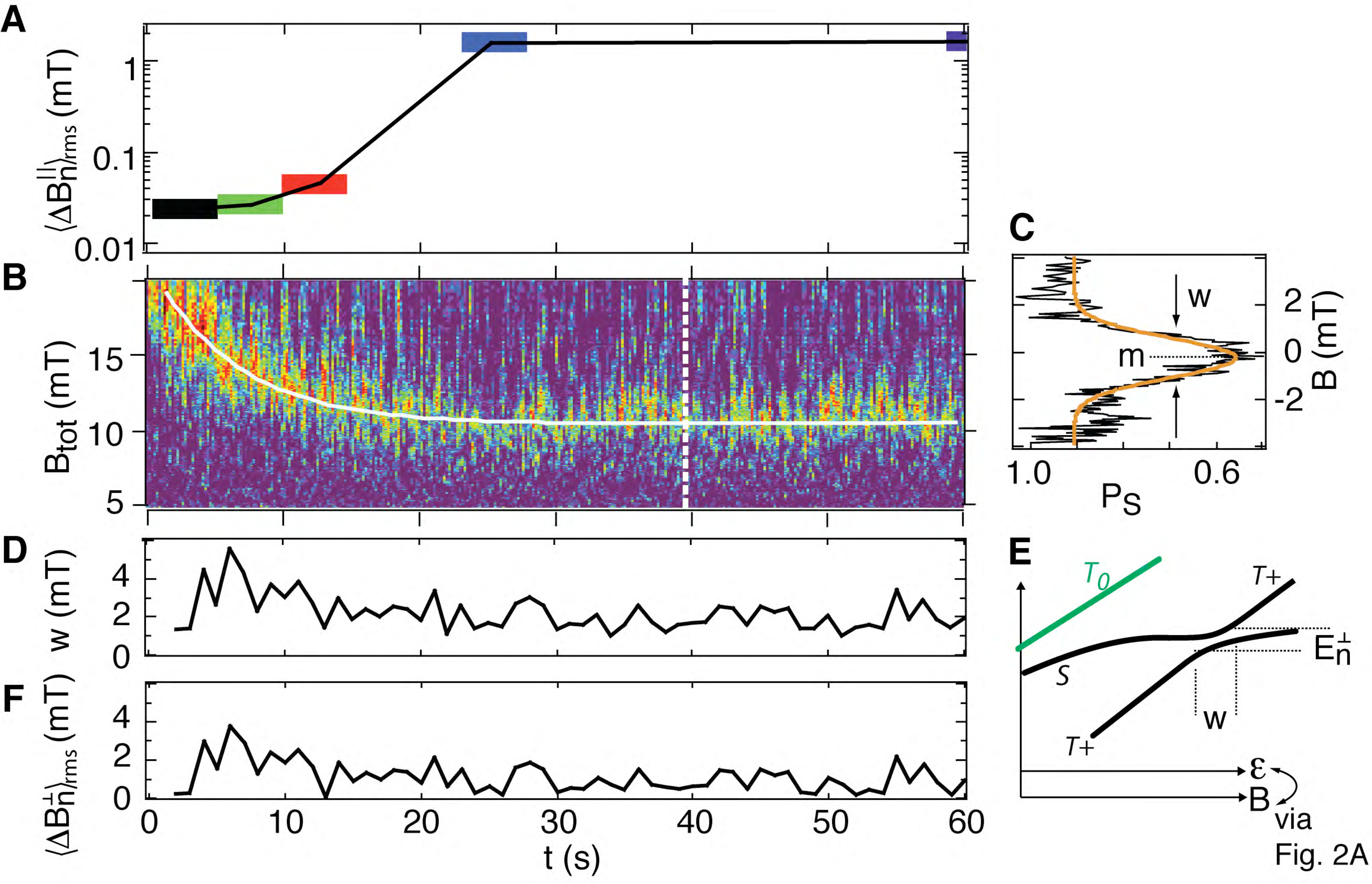}
\caption{
\noindent({\bf{A}}) Amplitude of fluctuating longitudinal Overhauser field, $\langle\dBpa\rangle_\mathrm{rms}$, extracted from $T_{2}^{*}$ values at the 5 s intervals in Fig.~3A. ({\bf{B}}) $S-T_{+}$ resonance probed immediately following the pump cycle. Position of the resonance yields $B_\mathrm{tot}$ and its intensity gives $P_{S}$ ($B_\mathrm{0}$ = 10 mT, $\tau_\mathrm{S} = 25$ ns). ({\bf{C}}) Slice from (B) at position marked by white dashed line, averaged for 1 s. For each slice a gaussian fit yields the mean position, $m$, and width, $w$, of the $S-T_{+}$ resonance, given in units of magnetic field via Fig.~2(A).  ({\bf{D}}) Resonance width $w$ as a function of time following the pumping cycle.  ({\bf{E}}) Schematic of the $S-T_{+}$ anticrossing showing how fluctuations of $E_\mathrm{n}^\mathrm{\perp}$ due to $\dBpe$ give the resonance a width. ({\bf{F}}) Fluctuations in $\dBpe$, as an rms strength $\langle\dBpe\rangle_\mathrm{rms}$ in a 1 s slice.}
\vspace{-0.5cm}
\end{center}
\end{figure}

The observation that an adiabatic electron-nuclear flip-flop cycle will suppress fluctuations of the nuclear field gradient has been investigated theoretically~\cite{ramon07,witzel07}. These models explain some but not all of the observed phenomenology described here, and it is fair to say that a complete physical picture of the effect has not yet emerged. Other nuclear preparation schemes arising from various hyperfine mechanisms, not directly related to the specific pump cycle investigated here, have also been addressed theoretically in the recent literature \cite{christ07,danon08}. 

Control of spin qubits in the presence of time-varying equilibrium Overhauser gradients require complex pulse sequences \cite{petta05} or control of sizable magnetic field gradients \cite{levy02,wu02}. Suppressing fluctuations of $\dBpa$ by a factor of order 100, as demonstrated here using nuclear state preparation, leads to an improvement in control fidelity of order  $10^4$, assuming typical control errors, which scale as $ (\Delta B_\mathrm{n}/ \dBpa)^2$ for low-frequency noise. We further anticipate generalizations of the present results using more than two confined spins that allow arbitrary gradients in nuclear fields to be created by active control of Overhauser coupling.

We thank  L. DiCarlo and A. C. Johnson for technical
 contributions and W. Coish, F. Koppens, D. Loss, and A. Yacoby for useful discussions. This work was supported in part by ARO/IARPA, DARPA, and NSF-NIRT (EIA-0210736). Research at  UCSB supported in part by QuEST, an NSF Center.

\end{document}